\documentclass{article}

\usepackage{amsfonts}
\usepackage{epsfig}

\begin{document}

\title{ Magnetohydrodynamics In The Context Of Nelson's Stochastic Mechanics}

\vspace{1cm}

\author{{D. Volchenkov$^{1}$  and  R. Lima$^{2}$}}

\vspace{0.5cm}

\date{\today}
\maketitle

\leftline{\small \texttt{ ${}^1$ Bielefeld-Bonn Stochastic Research Center (BiBoS),}}

\leftline{\small \texttt{ Bielefeld University, Postfach 100131, 33501, Bielefeld, Germany,}}

\leftline{\small \texttt{ Email: volchenk@physik.uni-bielefeld.de}}

\leftline{\small\texttt{ ${}^2$ Centre de Physique Theorique, CNRS Luminy Case 907,}}

\leftline{\small\texttt{ F-13288, Marseille, France,}}

\leftline{\small\texttt{ EMail: lima@cpt.univ-mrs.fr}}

\begin{abstract}
A simple generalization of the MHD model accounting for
the fluctuations of the configurations due to kinetic effects
in plasmas in short times small scales is considered. The velocity of conductive fluid and the magnetic field are considerd as the stochastic fields (or random trial trajectories) for which the classical MHD equations play the role of the mean field equations in the spirit of stochastic mechanics of E. Nelson.
\end{abstract}

\vspace{0.5cm}

\leftline{\textbf{ PACS codes: } 47.10ad, 47.27.E-, 47.27.ef}
\vspace{0.5cm}

\leftline{\textbf{ Keywords: }  }

\section{Introduction}
 \noindent

Theoretical investigations of  cross-field transports in the operating
ITER-FEAT (International Thermonuclear Experimental Reactor) calls for
an increasing confidence
in the modelling efforts that force one to search for the new principles of simulations.
The aim of our work is to provide a possible ground for
the optimization of existing numerical simulation algorithms for
the large scale simulations in hydrodynamics.

The set of equations which describe magnetohydrodynamics (MHD) are a
combination of the Navier-Stokes equations (\ref{MHD}.1) of fluid dynamics
and Maxwell's equations (\ref{MHD}.2) of electromagnetism,
\begin{equation}
\label{MHD}
\left\{
\begin{array}{lcr}
\dot{\bf{v}}+\nabla P +(v\cdot\nabla){\bf v}-(b\cdot\nabla){\bf b} &=& \nu \Delta \bf{v}\\
\dot{\bf{b}} + (v\cdot\nabla){\bf b} - (b\cdot\nabla){\bf v}& = & \lambda\Delta{\bf b},\\
\end{array}
\right.
\end{equation}
in which $\lambda =c^2/4\pi\sigma$ is the resistivity constant,
the inverse Prandtl-type constant. The equations (\ref{MHD})
describe the dynamics of electrically conducting fluids. The fluid
velocity ${\bf v}(x,t),$ $x\in \mathbb{R}^d$ is supposed to be
incompressible, so that the mass continuity equation for that is
reduced to the transversally condition $\nabla\cdot {\bf v}=0$.
The normalized magnetic field ${\bf b}(x,t)$ is
\begin{equation}
\label{magneticfield}
{\bf b}(x,t)\,=\,\frac{{\bf B}(x,t)}{\sqrt{4\pi\rho}}, \quad \nabla \cdot {\bf b}=0,
\end{equation}
in which ${\bf B}(x,t)$ is the magnetic induction and $\rho$ is the density of medium.
In addition to the usual hydrodynamical interaction presented in the
Navier-Stokes equation, in the moving medium, there is the Lorentz force
 \cite{Landau} exerted on charged particles in the electromagnetic field,
\begin{equation}
\label{Lorentz}
[\mathrm{curl}{\ }{\bf B}\times {\bf B}] =({B}\cdot\nabla){\bf B}-\nabla(B^2/2).
\end{equation}
  The first term in (\ref{Lorentz}) is amended to the hydrodynamical
  interaction while the second term
redefines the pressure field in the medium,
\begin{equation}
\label{pressure}
P(x,t)\to p(x,t)+\frac{B^2}{2}.
\end{equation}
The dynamical equation describing the evolution of magnetic field
follows from the simplest form of Ohm's
law \cite{Landau},
\begin{equation}
\label{Ohm}
{\bf j}(x,t)=\sigma\left(
{\bf E}(x,t)+\frac 1c \left[{\bf \varphi}(x,t)\times {\bf B}(x,t)\right],
\right)
\end{equation}
in which $\sigma$ is the conductivity, $c$ is the speed of light, $\bf{E}$
 and $\varphi$ are the electric field and the
electrostatic potential respectively, and the Maxwell equations neglecting
the displacement current.

The purely longitudinal contributions of the pressure gradient $\nabla P$
and of interactions
can be eliminated from (\ref{MHD}) by the applying of transverse projection,
\begin{equation}
\label{projector}
\mathrm{P}^{\bot}_{ij}=\delta_{ij}-\frac{k_ik_j}{k^2},
\end{equation}
if written in the Fourier space.

The equations (\ref{MHD}) are applicable to the plasma if it is
strongly collisional, so that the time scale of collisions is
shorter than the other characteristic times in the system, and the
particle distributions are therefore close to Maxwellian. When
this is not the case (for instance in fusion plasmas), we are
interested in smaller spatial scales, in which it may be necessary
to use a kinetic model which properly accounts for the
non-Maxwellian shape of the distribution function. However,
because MHD is very simple, and captures many of the important
properties of plasma dynamics, it is often qualitatively accurate,
and therefore while accounting for the possible kinetic effects in
plasmas we are nevertheless interested to stay within the general
framework of the MHD approach.

In the present paper, we investigate a simple generalization of
the MHD model (\ref{MHD}) modelling fluctuations of the
configurations $\{{\bf v},{\bf b}\}$ due to kinetic effects in
plasmas. In the proposed model, we suppose that $\{{\bf v},{\bf
b}\}$ are the stochastic fields for which (\ref{MHD}) plays the
role of the mean field equations. Recently, we have implemented
the similar approach for the Burgers and Navier-Stocks equations
in \cite{VolchenLima}.

\section{Stochastic dynamics as the Brownian motion}
\noindent

It is well known that many problems in stochastic dynamics can be
treated as a generalized Brownian motion ${\left\langle {\delta
\left( {u - u\left( {{\bf x},t} \right)} \right)} \right\rangle} _{\xi}
,$ in which the classical random field indicating the position of
a particle $u({\bf x},t)$ meets a Langevin equation,
\begin{equation}
\label{Langevin}
\dot {u}\left({x,t} \right)\, =\, K(u) + Q\left( u \right) + \xi ,
\end{equation}
where $\xi $ is the Gaussian distributed stochastic force
characterized by the correlation function
\[
D_{\xi}\,  =\,
{\left\langle \,\xi \xi\, \right\rangle}.
\]
Here the angular brackets
${\left\langle {\ldots} \right\rangle} _{\xi}  $ denote an average
position of particle with respect to the statistics of $\xi$.
$K(u)$ is the linear  differential operator, and
 $Q(u)$ is some $t$-local (independent of time derivatives)
 nonlinear term which depends on the position
$u({\bf x},t)$ and its spatial derivatives. Such a representation was a
key idea of the famous Martin-Siggia-Rose (MSR) formalism,
 \cite{Martin:1976}-\cite{DeDom}.

An elegant way to obtain the field theory representation of
stochastic dynamics is given by the functional integral
\begin{equation}
\label{1}
 {\left\langle {\delta \left( {u - u\left( {{\bf x},t}
\right)} \right)} \right\rangle} _{\xi}  \equiv \int {D\xi \exp\,
{\rm t}{\rm r}\left( { - {\frac{{1}}{{2}}}\xi D_{\xi}  \xi}
\right)\delta \left( {u - u\left( {{\bf x},t} \right)} \right)}
\end{equation}
where the tr-operation means the integration $\int {d{\bf x}\, dt} $ and
the summation over the discrete indices. The instantaneous
positions $u\left( {{\bf x},t} \right)$ meet the dynamical equation that
can be taken into account by the change of variables
\begin{equation}\label{2}
 \delta \left( {u - u\left( {{\bf x},t} \right)} \right) \to \delta
\left( {\dot {u}\left( {{\bf x},t} \right) - Q\left( {u} \right) - \xi}
\right)
\end{equation}
should the solution of dynamic equation exists and is unique. The
use of integral representation for the $\delta - $function in
(\ref{1}) transforms it into
\begin{equation}\label{3}
 \int {D\xi\, Du Du'\,\exp \,{\rm t}{\rm r}\left( { - {\frac{{1}}{{2}}}\xi
D_{\xi} \xi - u'\,\dot{u} + u'\,Q\left( {u} \right) + u'\,\xi} \right)}
\det M,
\end{equation}
in which $u'\left( {{\bf x},t} \right)$ is the auxiliary field that is
not inherent to the original model, but appears since we treat its
dynamics as a Brownian motion. The Jacobian $\det M$ relevant to
the change of variables (\ref{2}) is discussed later.

The Gaussian functional integral with respect to the stochastic
force $\xi$ in (\ref{3}) is calculated
\begin{equation}
\label{4}
\int \,{DuDu'} \,\exp S\left( u,u'\right)\,
\det M,
\end{equation}
in which
\begin{equation}
\label{S}
\quad S(u,u')\, = \,{\rm t}{\rm r}{\left[ { -
{\frac{{1}}{{2}}}u'D_{\xi}  u' - u'\dot {u} + u'Q\left( {u}
\right)} \right]}.
\end{equation}
By means of that all configurations of $\xi $ compatible with the
statistics are taken into account. The integral (\ref{3})
identifies the statistical averages ${\left\langle {\ldots}
\right\rangle} _{\xi} $ with the functional averages of weight
$\exp S$. The formal convergence requires the field $u$ to be real
and the field ${u}'$ to be purely imaginary.

The functional averages in (\ref{4}) can be represented by the
standard Feynman diagram series exactly matching (diagram by
diagram) the usual diagram series found by the direct iterations
of the Langevin equation averaged with respect to the random force
This fact justifies the use of functional integrals in stochastic
dynamics at least as a convenient language for the proper diagram
expansions.

The Jacobian $\det M$ in (\ref{3}) depends upon the nonlinearity
$Q\left( {u} \right)$. If $Q\left( {u} \right)$ does not depend
upon the time derivatives, all diagrams for $\det M$ are the
cycles of retarded lines $ \overleftarrow{\Delta}\propto \theta
\left( {t - t'} \right)$ and equal to zero excepting for the very
first term,
\begin{equation}
\label{6} \det M \,= \,const \cdot \exp {\rm tr}\Delta ,
\end{equation}
Then, the
convention is used for the Heaviside function of zero argument,
$\theta(0)=0$, so that $\det M = const$, \cite{Adzhemyan:1999}.

The functional averages computed with respect to the statistical
weight $\exp\, S$ can be expanded into the series of Feynman
diagrams drawn with the interaction vertices determined by the
nonlinearity $Q(u)$ and two propagators (lines) which have the
following analytical representations (in the Fourier space)
\begin{equation}
\label{8}
 \Delta_{uu'}\, =\,
\frac{1}{\left(- i\omega +
\tilde{K} \right)},\quad \Delta_{uu}\, =\,
\frac{D_\xi } {\left( \omega ^{2} + \tilde{K}^2 \right)},
\end{equation}
in which $\tilde{K}$ is the Fourier image of the linear part in the Eq.(\ref{Langevin}).
The inverse  Fourier transform of $\Delta_{uu'} $ shows that it is
retarded, $\Delta_{uu'} \propto \theta(t-t_0).$

Many dynamical systems are driven by the non-random external
forces. It is worth to mention that if one assumes  $\left| \xi
\right|\rightarrow 0$ (and consequently $D_\xi\rightarrow 0$), then the
Feynman diagram series is trivial since the propagator vanishes,
$\Delta_{uu} = 0$.
 However, the diagram series would be
recovered by means of regular external forcing \cite{VolchenLima} that gives
rise to a branching representation of stochastic dynamics.

\section{Probabilistic interpretations for the solutions of elliptic equations}
\noindent

The striking similarity between the Schr\"{o}dinger equation written for
 free particles and the
diffusion equation motivated the search for a stochastic
interpretation of the quantum mechanics.
The first attempt had been made by E. Schr\"{o}dinger himself
\cite{Shredinger} and accomplished
by J.C. Zambrini who derived the genuine Euclidean version of
quantum mechanics \cite{Zambrini}.

As a counter motion, E. Nelson \cite{Nelson66} had proposed a
 generalization of the theoretical scheme of classical mechanics
 known as {\it stochastic mechanics}. Classical deterministic trajectories are
 substituted by random trajectories of well defined stochastic processes.
 Under appropriate conditions, and for a large class of dynamical systems,
 the basic equations of
 stochastic mechanics show a surprising connection with the basic equations
 of quantum mechanics,
 \cite{Nelson67}-\cite{Nelson82}.
Therefore, stochastic mechanics gives an approach to quantization of dynamical
 systems, based on methods of probability theory and stochastic processes. The
 original formulation of stochastic mechanics rests on two basic hypothesis,
\cite{Blanchard}-\cite{Guerra95}.
The first assumes that the trajectories of the dynamical system are perturbed
 by an underlying Brownian motion.
The second is a particular form of the second principle of dynamics, where the
 classical acceleration is replaced by a suitable form of stochastic acceleration.
 Further developments of the theory show that the basic equation of stochastic
  mechanics can be derived from variational principles, in complete analogy with
   classical mechanics, based on the same classical action, but exploiting
   stochastically perturbed trajectories as trial trajectories \cite{Guerra95}.
 The basic equations of stochastic mechanics (the continuity equation and Madelung equation)
can be immediately connected with the Schroedinger equation of
quantum mechanics.
The entire operator structure of the quantum mechanical observables can be
 easily derived from
the general structure of stochastic mechanics.
 From this point of view, stochastic mechanics can be considered as a kind
 of probabilistic simulation
of quantum mechanics, \cite{Blanchard}.

Stochastic mechanics can be based on variational principles of Lagrangian type.
In order to obtain that it is necessary to generalize the action of classic mechanics
to the case where the trial trajectories belong to stochastic processes.
This program has been partially realized in \cite{Aldorvandi89}-\cite{Aldorvandi90}
for dynamical systems on curved manifolds.
This was the analog of the well known problem of writing the Feynman path integral
 for a quantum system on a curved manifold, so that all the results could be
 immediately translated into the language of Feynman path integrals.

Independently of quantum mechanics and quantum field theory, a
probabilistic interpretation for the solutions of linear elliptic
and parabolic equations with Cauchy and Dirichlet boundary
conditions had been proposed \cite{Courant}.  A stochastic process
had been defined for which the mean values of some functionals
coincide with the solution of the deterministic equations. The
problem of existence of such the probabilistic representations for
the certain classes of nonlinear equations has been studied
extensively by Dynkin \cite{Dynkin}. The probabilistic
representations of the Fourier transformed Navier-Stokes and
Burger's equations had been discussed in \cite{Sznitman} and later
extensively developed in the
 works of Oregon group \cite{Waymire}-\cite{Ossiander}. A stochastic
 representation for the Poisson-Vlasov equation
has been derived in \cite{Rui} recently. In all cases when the
appropriate stochastic process has been constructed,
its mean values is the solution of the mean field equation, but the
process itself always contains more information then
that of physical relevance in particular.

A model in which the classical deterministic trajectories
$u(x,t)$, $x\in \mathbb{R}^d$, satisfying the hydrodynamics
equations are substituted by the random trajectories of a
generalized Brownian motion
over the space of fluid velocity configurations
$u$
driven by the stochastic force $\xi$
is known as stochastic hydrodynamics \cite{MonYa}. Here,  $\xi $ is the Gaussian
distributed stochastic force characterized by the correlation function $D_{\xi}  =
{\left\langle {\xi \xi} \right\rangle},$ and the angular brackets
${\left\langle {\ldots} \right\rangle} _{\xi}  $ denote an average
velocity of particle with respect to the statistics of $\xi$.
In such a formulation, the above problem is equivalent to that one of
 Nelson's {\it stochastic mechanics}, \cite{Nelson67},\cite{Blanchard}.
The relevant variational principle leads to an action functional
of the Martin-Siggia-Rose (MSR) type \cite{Martin:1976}. The
MSR-theory of stochastic hydrodynamics has been formulated
independently by many authors, \cite{JanssenB}-\cite{Phythian}.
Diagram representations for
 the Green functions of stochastic hydrodynamics exactly reproduce
 the hydrodynamical
diagrams discussed by Wyld, \cite{Wyld}. In general, the Green
functions of stochastic hydrodynamics
diverge for very large moments and therefore require the
ultraviolet renormalization that has been
discussed in details in \cite{Adzhemyan:1999}.

In our previous paper \cite{VolchenLima}, we have pointed that the diagram
 representations for
the Green's functions in hydrodynamics is still nontrivial if one
considers a regular external forcing, instead of random one. In particular, we have
studied the $\delta(x-x_0)\delta(t-t_0)$ external forcing corresponding to
 the Cauchy problem of the Navier-Stokes equation supplied
with an integrable initial condition.
Each Feynman graph in the diagram series equals
to an average over a forest of multiplicative branching binary
trees (of the certain topological structure) implemented in
\cite{Waymire}-\cite{Ossiander}. The branching representations for the Green function
of Cauchy problem establishes the direct relation between Nelson's stochastic
 mechanics \cite{Nelson67} and
the  probabilistic interpretations for the solutions of nonlinear equations with Cauchy
 and Dirichlet boundary conditions studied in \cite{Dynkin}-\cite{Ossiander}.
It is important to note that in contrast to the MSR theory \cite{Adzhemyan:1999},
the diagrams of branching representations for the hydrodynamics equations
 \cite{VolchenLima} do not diverge,
but consitute a regular expansion starting from the standard diffusion kernel.
Diagram contributions represent the consequent bifurcations of media resulting in the
cascade of consequent partitions of moments,
$\mathbf{k}=\mathbf{q} + (\mathbf{k} - \mathbf{q})$. The magnitude
of relevant corrections to the standard diffusion spectrum tends to zero as
$t\rightarrow t_0$, and the saddle-points (instanton) analysis can
be then applied to study the "large order" asymptotic contributions \cite{VolchenLima}.
The calculations have shown that the asymptotic coefficients demonstrate
the factorial growth like for the most of models in quantum field
theory. The asymptotic series for the Green function can be
summarized by means of the Borel procedure. In the
limit $t\to t_0,$ the corrections to the diffusion kernel have the
closed analytical form.

\section{Cauchy problem for the MHD equations and its stochastic mechanics formulation}
\noindent

The Cauchy problem for the MHD equations,
\begin{equation}
\label{MHDCauchy}
\left\{
\begin{array}{lcl}
\dot{v}_i+\partial_j\left(v_iv_j-b_ib_j\right)-\nu \partial^2 v_i&= & \delta(t-t_0)\delta(x-x_0), \\
\dot{b}_i+\partial_j\left(v_jb_i-b_jv_i\right)- \lambda \partial^2 b_i &= &\delta(t-t'_0)\delta(x-x'_0),
\end{array}
\right.
\end{equation}
is supplied with the localized integrable initial conditions ${\bf v}_0={\bf v}(x,0)$ and
${\bf b}_0={\bf b}(x,0)$.
It is possible to construct a stochastic counterpart of MHD (\ref{MHDCauchy})
by adding a Gaussian
distributed random force into it. This approach leads to the stochastic
magnetohydrodynamics
\cite{Fournier},\cite{Adzhemyan:1999},\cite{Habilitation}
 developed in order to study the inertial range scaling laws and different
 regimes of large-scale
asymptotic behavior. Alternatively, in the framework of stochastic
interpretation, we consider a stochastic model in which the
nonlinear dynamical equations (\ref{MHDCauchy}) play the role of
{\it mean field} equations, so that their solutions satisfying the
given initial conditions play the role of the observables. Instead
of the classical deterministic fields ${\bf b}(x,t)$ and ${\bf
v}(x,t)$, we study their stochastic trial analogs,
$\widetilde{{\bf b}}(x,t)$ and  $\widetilde{{\bf v}}(x,t)$. Then
the Green functions $G_\phi$, $\widetilde{ \phi}\, = \,
\{\widetilde{\bf v},\widetilde{\bf b}\}$ of the original Cauchy
problem (\ref{MHDCauchy}) can be represented by the functional
averages,
\begin{equation}
\label{G_v}
G_\phi(x,t;x_0,x'_0;t_0,t'_0)\,=\,\frac{\int {\mathcal D}\!\Phi
{\ }\widetilde{\phi}(x,t)\exp S(\Phi)}{\int {\mathcal D}\!\Phi \,\exp S_0(\Phi)},
\end{equation}
over all possible configurations $\widetilde{{\bf v}}(x,t)$ and $\widetilde{{\bf b}}(x,t)$
 such that
their expectation values satisfy (\ref{MHDCauchy}) with the given initial conditions.
 Here, $\Phi$ are the functional arguments of the   action functional
$S(\Phi)$ such that (\ref{MHDCauchy}) are its saddle-point equations.
Its "quadratic" part, $S_0(\Phi)$, corresponds to the linearized equations
of MHD which play the role of an interaction free theory.
We discuss the measure of functional integration
${\mathcal D}\!\Phi$ after we consider $S(\Phi)$. If we introduce the
 auxiliary fields $\widetilde{{\bf v'}}(x,t)$  and
$\widetilde{{\bf b'}}(x,t)$, then, up to an inessential constant factor, the relevant action
functional reads as following:
 \begin{equation}
\label{action}
\begin{array}{lcl}
S(\widetilde{v},\widetilde{v'},\widetilde{b},\widetilde{b'}) & =& \widetilde{v'}(x_0,t_0)+ \widetilde{b'}(x'_0,t'_0)\\
& &-\mathrm{tr}\left[
\widetilde{v'}_i\dot{\widetilde{v}}_i + \nu\widetilde{v'}_i\partial_j(g_1\widetilde{v}_i\widetilde{v}_j-
g_2\widetilde{b}_i\widetilde{b}_j)-\nu \widetilde{v'}_i\Delta \widetilde{v}_i\right] \\
& & - \mathrm{tr}\left[\widetilde{b'}_i\dot{\widetilde{b}}_i
+\lambda\widetilde{b'}_i\partial_j(g_3\widetilde{v}_j\widetilde{b}_i-g_4
\widetilde{b}_j\widetilde{v}_i)-\lambda\widetilde{b'}_i
 \Delta \widetilde{b}_i
\right]
\end{array}
\end{equation}
where as in (\ref{S}) the $\mathrm{tr}$-operator means the integration $\int\! dx\,dt$ and summation over the discrete indices.
The quadratic part of (\ref{action}) has the form
\begin{equation}
\label{S_quadratic}
S_0(\widetilde{v},\widetilde{v'},\widetilde{b},\widetilde{b'}) \, =\,
-\mathrm{tr}\left[
\widetilde{v'}_i\dot{\widetilde{v}}_i +\widetilde{b'}_i\dot{\widetilde{b}}_i -\nu \widetilde{v'}_i\Delta \widetilde{v}_i-\lambda\widetilde{b'}_i
 \Delta \widetilde{b}_i\right].
\end{equation}
The action functionals with ultra-local terms like those presented
in (\ref{action}), $\widetilde{v'}(x_0,t_0)$ and
$\widetilde{b'}(x_0,t_0)$, had been studied in
\cite{Symanzik:1981}. In order to obtain the formal expansion
parameters in the perturbation theory for the above action
functional, we have inserted four coupling constants,
$g_{1,2}\nu\equiv 1$, and $g_{3,4}\lambda \equiv 1$, in front of
the relevant interaction terms in (\ref{action}). Despite their
physical dimensions are $[g_{1,2}]=-[\nu]$ and
$[g_{3,4}]=-[\lambda]$, their convectional dimensions would be
different. If we put formally $g_{1,2,3,4}=0$ in (\ref{action}),
the action functional $S(\Phi)$ turns into $S_0(\Phi)$, free of
interactions. The correct normalization of integral in (\ref{G_v})
requires that the measure of functional integration
$\mathcal{D}\!\Phi$ be normalized to the volume of orbits in the
free theory (pure diffusion processes),
\begin{equation}
\label{measure}
\mathcal{D}\!\Phi=\frac{\mathcal{D}\!\widetilde{\bf v}\mathcal{D}\!\widetilde{\bf v'}\mathcal{D}\!\widetilde{\bf b}\mathcal{D}\!\widetilde{\bf b'}}{Z}, \quad Z\equiv \int \mathcal{D}\!\Phi \exp S_0(\Phi).
\end{equation}
Moreover, it is obvious that the results of functional averages
(\ref{G_v}) do not change along the set of orbits in the
configuration space related by any symmetry transformation of the
action (\ref{action}). The functional integral (\ref{G_v}) itself
is proportional to the volume of such orbits. The functional
(\ref{action}) possess the Galilean invariance:
\begin{equation}
\label{Galilean}
\begin{array}{lcl}
\widetilde{\bf v}({\bf x},t)& =& \widetilde{\bf v}({\bf x}+{\bf s}(t),t)-{\bf u}(t),\\
\widetilde{\bf v'}({\bf x},t)& =& \widetilde{\bf v'}({\bf x}+{\bf s}(t),t),\\
\widetilde{\bf b}({\bf x},t)& =& \widetilde{\bf b}({\bf x}+{\bf s}(t),t),\\
\widetilde{\bf b'}({\bf x},t)& =& \widetilde{\bf b'}({\bf x}+{\bf s}(t),t),
\end{array}
\end{equation}
where an integrable function ${\bf u}(t)$ is a parameter of
transformation (the velocity of the frame of reference),
and its integral ${\bf s}(t)=\int^t_{-\infty}\!{\bf u}(t')dt'$.

The auxiliary fields introduced in (\ref{action}) were not inherent
 to the original physical model, but appear
since we treat its dynamics as a Brownian motion.
While the first two saddle point equations,
\begin{equation}
\label{saddle1}
\frac{\delta S}{\delta \widetilde{v'}} =0, \quad \frac{\delta S}{\delta \widetilde{b'}}=0,
\end{equation}
recover the original Cauchy problem (\ref{MHDCauchy})
(in case $g_{1,2}\nu=1$, $g_{3,4}\lambda=1$), another pair,
\begin{equation}
\label{saddle2}
\frac{\delta S}{\delta \widetilde{v}} =0, \quad \frac{\delta S}{\delta \widetilde{b}}=0,
\end{equation}
describes the dynamics of auxiliary fields:
\begin{equation}
\label{auxil}
\left\{
\begin{array}{lcl}
\dot{\widetilde{ v'}}_i+\partial_j(g_1\nu\widetilde{ v'}_i\widetilde{ v}_j-g_4\lambda\widetilde{ b'}_i\widetilde{ b}_j)&= &
 -\nu \Delta \widetilde{ v'}_i, \\
\dot{\widetilde{ b'}}_i+\partial_j(g_3\lambda\widetilde{ b'}_i\widetilde{ v}_j-g_2\nu\widetilde{ v'}_i\widetilde{ b}_j)&= &
-\lambda \Delta \widetilde{ b'}_i.
\end{array}
\right.
\end{equation}
The above equations are characterized by the negative dissipations
since $\nu,\lambda>0$. Therefore, physically relevant solutions
have
 to satisfy ${\bf \widetilde{v'}}(t>0)=0$ and ${\bf\widetilde{ b'}}(t>0)$.
The equations (\ref{MHDCauchy},\ref{auxil}) give the time
evolution of the infinitesimal characteristics of the diffusion
processes in MHD and therefore determines them. It is important to
mention the striking similarity between the equations
(\ref{MHDCauchy}, \ref{auxil}) and the coupled system of nonlinear
equations for the osmotic velocity and the current velocity of
stochastically driven Brownian motion,
\cite{CarlenA}-\cite{ZhengB}.

We conclude the current section with a remark that the action
 functional (\ref{action}) can be
derived form the standard action functional of the MSR-type for
 stochastic MHD \cite{Fournier},\cite{Adzhemyan:1999}
with the ultra-local interaction terms ${\bf v'}(x_0,t_0)$
and ${\bf b'}(x'_0,t'_0)$ added, if one put
the stochastic forcing to zero, $|\xi|\to 0$.

\section{Diagram technique for stochastic MHD. The absence of ultraviolet divergences}
\noindent

The formal functional averages (\ref{G_v}) computed with respect
to the statistical weights $\exp S(\Phi)$ are interpreted as the
infinite diagram series,
\begin{equation}
\label{diagrams}
G_\phi\left(x,t;x_0,t_0;x'_0,t'_0\right)\,=\,\frac 1{k_1!k_2!k_3!k_4!}\sum_{
\{k_1+k_2+k_3+k_4\,=\,0\}}^{\{k_1+k_2+k_3+k_4\,=\,\infty\}}
G^{(k_1,k_2,k_3,k_4)}_\phi g^{k_1}_1g^{k_2}_2g^{k_3}_3g^{k_4}_4,
\end{equation}
where the sum is taken over all
positive integer solutions of the equation
\begin{equation}
\label{integer}
k_1+k_2+k_3+k_4\,=\,N, \quad k_i,\,N\,\in \,\mathbb{Z}_+.
\end{equation}
The series (\ref{diagrams}) starts from the standard diffusion
kernel in the coupling constants representing the possible
physical interactions in MHD:  the hydrodynamic dragging ($g_1$),
the Lorenz force ($g_2$), the convection ($g_3$), and the
stretching ($g_4$). In view of that, the main tasks of the
stochastic approach to MHD is the computation of coefficients
$G^{(k_1,k_2,k_3,k_4)}_\phi$, and the estimation of the asymptotic
properties of the diagram expansions
 (\ref{diagrams}).

In general, the computation of $G^{(k_1,k_2,k_3,k_4)}_\phi$ in
 (\ref{diagrams}) is a very difficult problem.
However, if the functional averages (\ref{measure}) are taken over
the Gaussian distributed fields, one can apply Wick's theorem
\cite{Zinn}-\cite{Bog} to facilitate calculations. In accordance
 to Wick's theorem, the functional averages  (\ref{G_v})
equal to the sums of all complete systems of "pairings" of the
interaction operators. If we denote operators as nodes and the
pairings between them as edges, then each system of "pairings"
representing the particular average can be visualized by a graph
called a Feynman diagram. The use of Feynman diagrams  helps to
make the computations visual and clarifies their similarity with
the problems of symbolic dynamics \cite{Markus}.

\begin{figure}[ht]
\begin{center}
 \epsfig{file=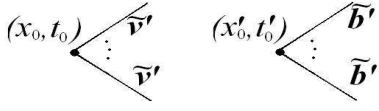,angle=0,width=5cm,height=1.5cm}
\end{center}
\caption{The ultra-local interaction vertices in stochastic
representation of magnetohydrodynamics.}
\label{Fig00}
\end{figure}

Any Feynman diagram for the stochastic representation of MHD  contains one of two
ultralocal interaction terms (see Fig.~\ref{Fig00}),
with any number of either $\widetilde{\bf b'}-$ or $\widetilde{\bf
v'}-$tails located at the points $(x'_0,t'_0)$ and $(x_0,t_0)$
consequently. They reveal the dependence of the responde functions
$G_v$ and $G_b$ in MHD upon the initial conditions for the
velocity field at $(x_0,t_0)$ and for the magnetic field at
$(x'_0,t'_0)$. In quantum field theory, the vertices shown on
(\ref{Fig00}) are called the composite operators, \cite{Zinn}.

\begin{figure}[ht]
\begin{center}
 \epsfig{file=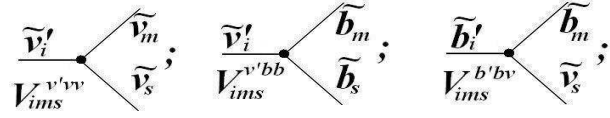,angle=0,width=8cm,height=1.5cm}
\end{center}
\caption{The interaction vertices in stochastic representation of
 magnetohydrodynamics. The vertex factors $V^{\phi'\phi\phi}$ in the
  Fourier representation are given in (\ref{letters}). }
\label{Fig01}
\end{figure}

\begin{figure}[ht]
 \epsfig{file=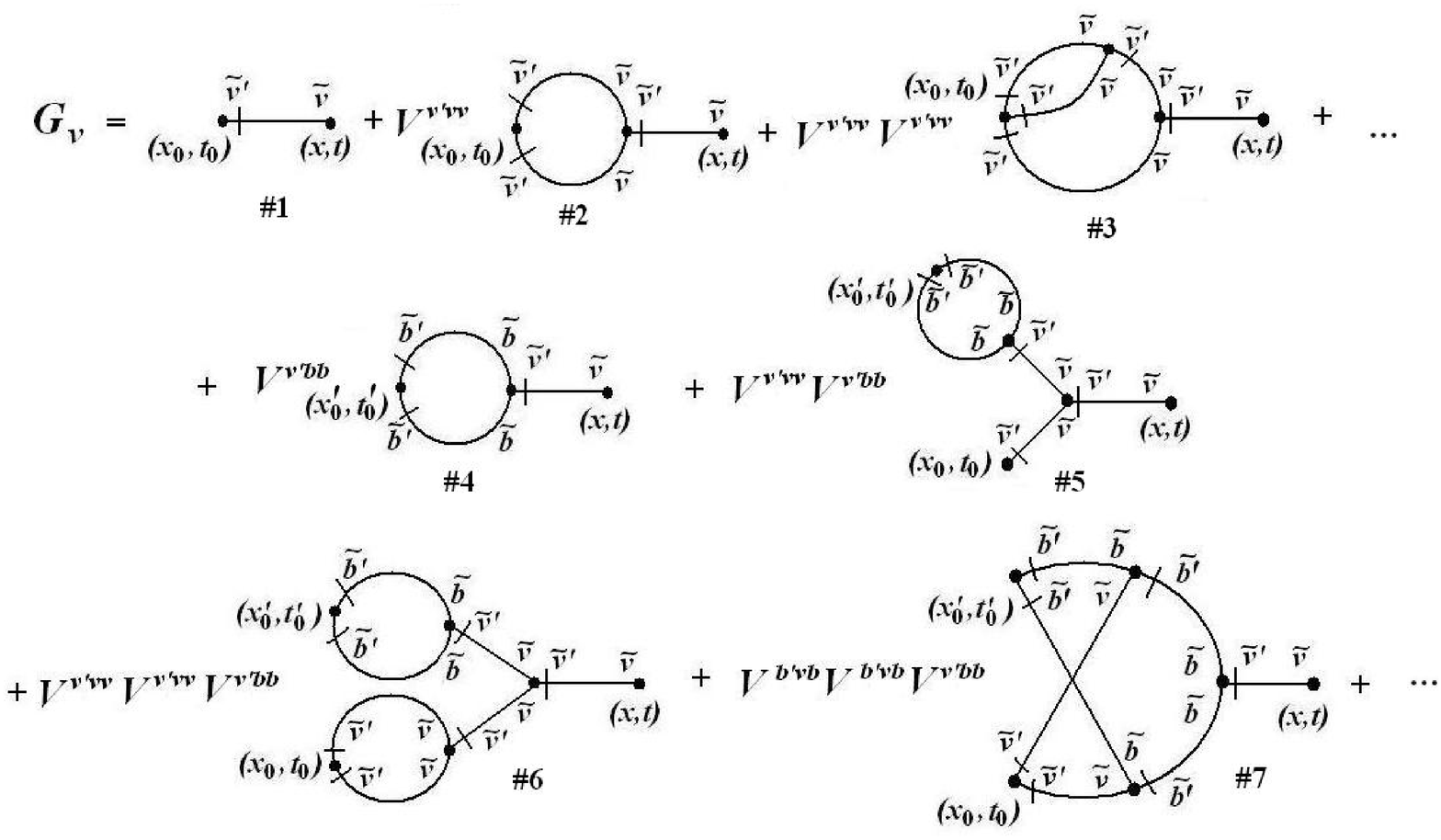,angle=0,width=12cm,height=7cm}
\caption{\small The first diagrams for the inverse exact responde
function of the fluid velocity in MHD. The slash marks auxiliary
fields.  The first diagram corresponds to the responde function of
the linearized problem. The second diagram expresses the first
order correction to the responde function due to the hydrodynamic
drag. The third diagram stands for the second order correction due
to the hydrodynamic drag. The forth diagram stands for the first
order correction due to the Lorenz force. The forthcoming diagrams
are related to the corrections risen by the combined effect of
hydrodynamic and magnetic field fluctuations.} \label{Fig1}
\end{figure}

\begin{figure}[ht]
 \epsfig{file=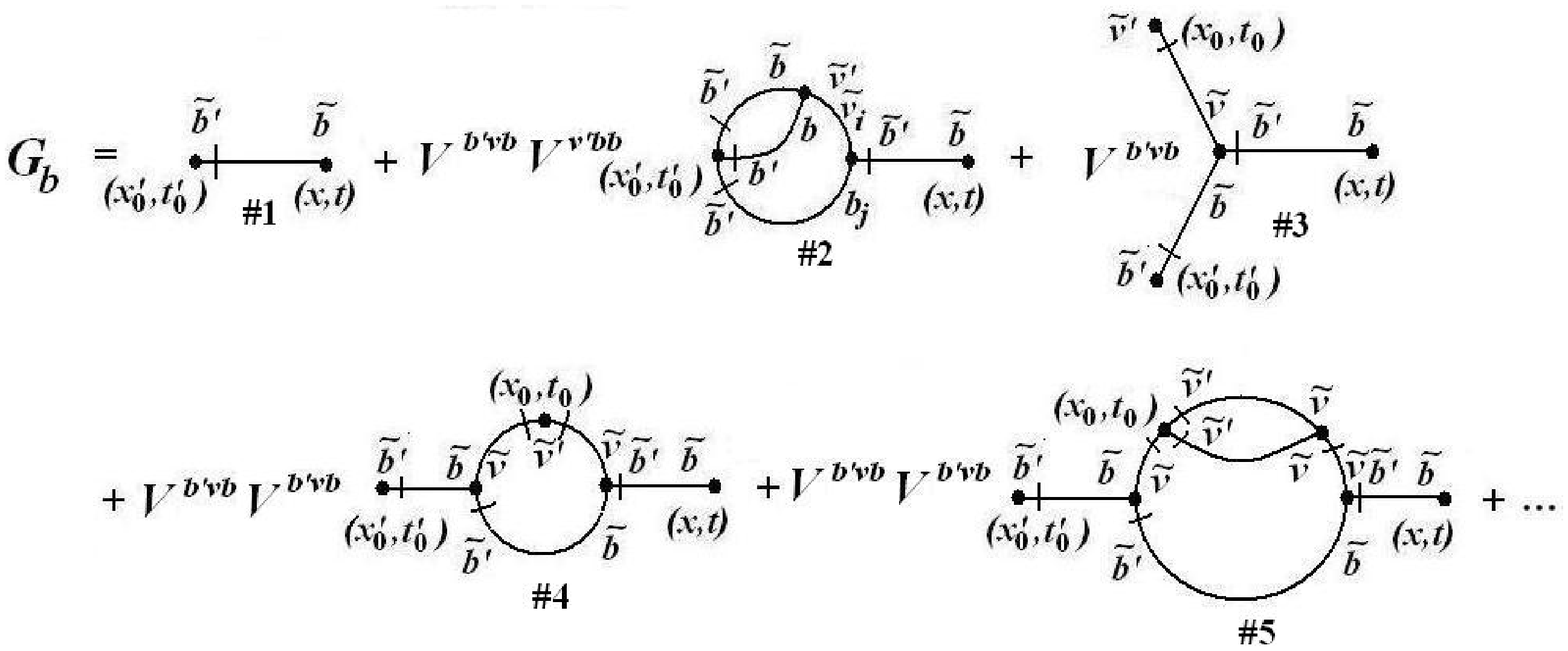,angle=0,width=12cm,height=5cm}
\caption{\small The first diagrams for the inverse exact responde
function of the magnetic field in MHD.
  The first diagram corresponds to the responde function of the linearized
  problem, $R_{ij}^{b'b}$, in (\ref{kernels}).
The second diagram gives the second order (in the coupling
constants) one-point correction due to fluctuations of the
magnetic field.  The third diagram stands for the first order
two-points contribution. The forth diagram describes the the
combined effect of the convection and the stretching. The fifth
diagram gives the second order two-points correction due to the
combined effect of hydrodynamic drag, convection, and stretching.}
\label{Fig2}
\end{figure}

 A diagram consists of the
"interaction vertices" (see Fig.~\ref{Fig01}),
with the factors $V^{v'vv}_{ims}$, $V^{v'bb}_{ims}$, and $V^{b'bv}_{ims}$,
which can be written in
Fourier space as
\[
V^{v'vv}_{ims}=\frac {g_1\nu}2 (i\delta_{im}k_s+i\delta_{is}k_m),\quad
V^{v'bb}_{ims}=-\frac {g_2\nu}2 (i\delta_{im}k_s+i\delta_{is}k_m),
\]
\begin{equation}
\label{letters}
V^{b'bv}_{ims}=\lambda(ig_4\delta_{is}k_m-ig_3\delta_{im}k_s).
\end{equation}
The  $\widetilde{\bf b'}-$ or $\widetilde{\bf v'}$-tails in interaction
vertices (see Fig.~\ref{Fig01}) are to be
connected with the $\widetilde{\bf b}-$ or $\widetilde{\bf v}$-tails by means of edges
(the response functions of the linearized MHD equations):
\begin{equation}
\label{edges}
R_{ij}^{v'v}=\frac{\mathrm{P}^\bot_{ij}(k)}{-i\omega+\nu k^2}, \quad
R_{ij}^{b'b}=\frac{\mathrm{P}^\bot_{ij}(k)}{-i\omega+\lambda k^2}.
\end{equation}
in which the transverse projector $\mathrm{P}^\bot_{ij}(k)$ is
given in (\ref{projector}). The inverse Fourier transform with
respect to the frequency $\omega$ in (\ref{edges}) reveals that
the response functions of linearized problem are retarded,
\begin{equation}
\label{kernels}
R_{ij}^{v'v}= H(t-t_0)\mathrm{P}^\bot_{ij}(k)e^{-\nu k^2(t-t_0)},
\quad
R_{ij}^{b'b}= H(t-t_0)\mathrm{P}^\bot_{ij}(k)e^{-\lambda k^2(t-t_0)},
\end{equation}
where $H(t)$ is the Heaviside function supplied by the convention
$H(0)=0$. It is required by the  the {\it casualty} principle that
the time arguments correspondent to auxiliary fields
$\widetilde{\bf v'}$ and $\widetilde{\bf b'}$  always precedes the
time arguments of $\widetilde{\bf v}$ and $\widetilde{\bf b}$. Let
us note that the inverse Fourier transforms of response functions
(\ref{kernels}) correspond to the standard diffusion kernels, (in
$d$-dimensional space),
\begin{equation}
\label{gaussian} \Delta \left( x - x_0 ,t - t_0\right) =
\frac{\mathrm{e}^{- \frac{(x - x_0)^2}{4\nu (t-t_0)}}}{\sqrt{\left(4\pi\nu(t-t_0)\right)^{d}}}.
\end{equation}
Similarly to quantum electrodynamics where each Feynman diagram
represents a certain process of
physical interactions between elementary particles, any graph in stochastic mechanics
 drawn with the interaction vertices
 Figs.~\ref{Fig00},\ref{Fig01} connected by
the edges (\ref{edges}) corresponds to a certain process of
interactions between different magnetic and hydrodynamic modes
coming along with the cascades of
consequent partitions of moments, ${\bf k}={\bf q}+({\bf k-q})$.
 The first diagrams relevant to the inverse Green functions (exact response functions) of MHD
are shown in Fig.~\ref{Fig1} and Fig.~\ref{Fig2} consequently.
Each diagram in expansions corresponds to a certain analytical
integral expression quantifying the contribution of the relevant
interactions into the entire behavior (they are given in the
Appendix A). In general case, all integrals with an odd number of
factors $k_i$ equals zero as a consequence of space isotropy.

The diagram expansion for the Green function would have a definite
physical meaning if it converges. The standard analysis of
ultraviolet divergences of graphs is based on the counting of
relevant canonical dimensions. Dynamical models have two scales,
the time scale $T$ and the length scale $L$, consequently the
physical dimension of any quantity $F$ can be defined as ${\left[
{F} \right]} = L^{ - d_{F}^{k}} T^{ - d_{F}^{\omega} } ,$ in which
$d_{F}^{k} $ and $d_{F}^{\omega}  $ are the momentum and frequency
dimensions of $F$. In diffusion models, these dimensions are
always related to each other since
 $\partial _{t} \sim \partial^2_x $ in the diffusion equation that allows
us to introduce a combined canonical dimension, $d_{F} =
d_{F}^{\omega}  + 2d_{F}^{k} $. One can check out that each term
in (\ref{action}) is dimensionless if the following relations hold:
$d_{\phi'} = 0,d_{u} = d,d_{\nu,\lambda}  = - 2 + 2 \cdot 1 = 0,$ and $d_{g}
= 2 - \left( {d + 1} \right).$ The field theory (\ref{action}) is
logarithmic (the conventional dimension of the coupling constant
$d_{g_i} = 0$) for the Burgers equation $\left( {d = 1} \right)$,
while in two dimensions $d_{g_i} = - 1{\rm ,}$ and $d_{g_i} = - 2$ for
$d = 3$ (the NS equation). Thus, in the infrared region (small
moments, large scales) the diagram series in $g_i$ define just the
corrections to the diffusion kernel as $d \ge 2$. However, in the
case of Burgers equation, all diagrams look equally essential in
large scales.

The diagrams diverge in the ultraviolet region (large moments,
small scales) if their canonical dimension
\[
d_{\Gamma}  = - d_{\phi} N_{\phi} - d_{\phi'} N_{\phi'} \ge 0
\]
where $N_{\phi} $ and $N_{\phi'} $ are the numbers of corresponding
external legs in the graph $\Gamma$. For the Green function
(\ref{G_v}), we have $N_{\phi'} = 0$ and $N_{\phi} = 1$. Therefore, there
is no ultraviolet divergent graphs in the diagram series
(\ref{diagrams}). At first glance, it seems that any graph having
no external $\phi - $legs $\left( {N_{\phi} = 0} \right)$ and any number
of auxiliary fields $\phi'$ $\left( {N_{\phi'} > 0} \right)$ should
diverge since $d_{\phi'} = 0$. However, such a graph is also
convergent in small scales because of the derivatives in the vertex factors $V(k)$ which
are always taken outside the graph onto the external $\phi' - $legs
that effectively reduces its canonical dimension to ${d}'_{\Gamma}
= d_{\Gamma}  - N_{\phi'} < 0{\rm .}$ Therefore, the field theory with the action functional
(\ref{action}) has no ultraviolet divergences and does not need a renormalization.

Each Feynman diagram in (\ref{diagrams}) corresponds to a certain
magnetohydrodynamic process. For instance, the very first diagrams
displayed in Figs.~\ref{Fig1},\ref{Fig2} present the solutions of
diffusion equations (in $d$-dimensional space). They describe the
simple viscous dissipation of a hydrodynamic vortex with no
bifurcations and the motion of magnetic field through the fluid,
following a diffusion law with the resistivity of the plasma
$\lambda$ serving as a diffusion constant.

 The second graph ($\#2$) in Fig.~\ref{Fig1} corresponds to a
bifurcation characterized by the
  twofold splitting of the moment, ${\bf k}={\bf q} + ({\bf k}-{\bf q})$.
Under the spatial Fourier
transformation, it is equivalent to the following analytic
expression:
\begin{equation}
\label{15}
\begin{array}{lcl}
\Gamma _{1} \left( {{\rm {\bf k}},t} \right) &= & -
g{\int\limits_{0}^{\infty} {d{t}'}} \, {\bf k}\cdot\,
\Delta \left( {{\rm {\bf k}},t - {t}'} \right) \\
&  &\int
{{\frac{{d{\rm {\bf q}}}}{{\left( {2\pi} \right)^{d}}}}} \Delta
\left( {{\rm {\bf q}},{t}'} \right)v_{0} \left( {{\rm {\bf q}}}
\right)\Delta \left( {{\rm {\bf k}} - {\rm {\bf q}},{t}'}
\right)v_{0} \left( {{\rm {\bf k}} - {\rm {\bf q}}} \right),
\end{array}
\end{equation}
where $\Delta \left( {{\rm {\bf k}},t} \right)$ is the spatial
Fourier transform of the diffusion kernel and ${\bf v}_{0} \left(
{{\rm {\bf k}}} \right)$ is the Fourier spectrum of initial
condition. It is worth to mention that the diagram expansion for
the Green function can be also discussed for the function defined
on a finite domain supplied with  periodic boundary conditions.
In the latter case, it has a discrete set of harmonics, and the
integral (\ref{15}) turns into sums.

The time integration in (\ref{15}) can be performed easily,
\begin{equation}
\label{16}
\begin{array}{lcl}
\Gamma _{1} & = & - g\, {\bf k}\cdot\exp
\left( { - \nu k^{2}\left( {t - t_{0}}  \right)}
\right) \\
& \times & {\int_{{\rm {\bf q}} \cdot {\rm {\bf k}} > q^{2}}
{{\frac{{d{\rm {\bf q}}}}{{\left( {2\pi} \right)^{d}}}}{\left[
{v_{0} \left( {{\rm {\bf q}}} \right)v_{0} \left( {{\rm {\bf k}} -
{\rm {\bf q}}} \right){\frac{{\exp ( - 2\nu \left( {{\rm {\bf q}}
\cdot {\rm {\bf k}} - q^{2}} \right)t_{0} )}}{{2\nu \left( {{\rm
{\bf q}} \cdot {\rm {\bf k}} - q^{2}} \right)}}}} \right]}}} .
\end{array}
\end{equation}
The singularities in $\Gamma _{1} $ appear at ${\rm {\bf q}} = 0$
and ${\rm {\bf q}} = {\rm {\bf k}}$, when the vortex does not
bifurcate. The remaining momentum integral in (\ref{16}) can be
interpreted as an expectation value,
\begin{equation}
\label{17} \wp _{1} \left( {{\rm {\bf k}}} \right) =
{\int\limits_{t_{0}} ^{\infty} {d\tau} } {\int_{} {{\frac{{d{\rm
{\bf q}}}}{{\left( {2\pi} \right)^{d}}}}{\left[ {v_{0} \left(
{{\rm {\bf q}}} \right)v_{0} \left( {{\rm {\bf k}} - {\rm {\bf
q}}} \right)\exp \left( { - 2\nu \left( {{\rm {\bf q}} \cdot {\rm
{\bf k}} - q^{2}} \right)\tau}  \right)} \right]}}} ,
\end{equation}
over the Poisson process of vortex bifurcation at momentum ${\rm
{\bf k}}$. Bifurcation of vortexes is the Poisson stochastic
process developing with time \cite{VolchenLima}. Further diagrams
shown on Figs.~\ref{Fig1},\ref{Fig2} represent more complex
magnetohydrodynamic processes.

\section{The multiplicative factors of Feynman diagrams in MHD}
\label{Sec:factors}
\noindent

One of fascinating features of the proposed stochastic approach to
MHD is that all physically admissible behaviors of the MHD system
are encoded by the certain integer solutions of the
Eq.(\ref{integer}). It is clear that not all such the solutions
are equally contribute to the diagrammatic series
(\ref{diagrams}). While drawing diagrams admissible with respect
to the "grammar" prescribed by the standard Feynman rules
discussed in the previous section, one can see that only some
particular combinations of coupling constants can appear as the
factors before diagrams for the Green functions.

In order to find these factors, we use the standard methods of
graph theory. Indeed, each diagram constitutes a graph, in which
every node representing one of four physical  interactions is
proportional to the relevant coupling constant.  We start drawing
a diagram from the point $(x,t)$ by adding vertices and connecting
them in a way admissible with respect to the Feynman rules. At any
step, the tails representing the fields $\widetilde{\bf v}$ and
$\widetilde{\bf b}$ can be fused together at the vertices of the
ultra-local interactions (the composite operators) $\widetilde{\bf
v'}^m(x_0,t_0)$ and $\widetilde{\bf b'}^m(x'_0,t'_0)$, $m\geq 1$.
It is convenient to count the powers of coupling constants in the
multiplicative factors before diagrams of perturbation theory with
the use of the "grammar matrix" $\mathbb{G}$ given in the Appendix
\ref{Append:B}.

The grammar matrix is the weighted  connectivity matrix, which
expresses the fact that  the tails $\phi$, $\phi'$ belonging to
the interaction vertices $V^{v'vv}_{ims}$, $V^{v'bb}_{ims}$, and
$V^{b'bv}_{ims}$ (being proportional to $g_1$, $g_2$, $g_3$, and
$g_4$) can be connected to each other accordingly to the Feynman
rules only by means of the propagators (\ref{edges}). For example,
the first row of the grammar matrix $\mathbb{G}$  shows that the
$\widetilde{\bf v'}-$tail belonging to the vertex $V^{v'vv}_{ims}$
may be connected within a Feynman graph either to the
$\widetilde{\bf v}-$tails in the similar interaction vertex
$V^{v'vv}_{ims}$ or to the $\widetilde{\bf v}-$tails in the
interaction vertex $V^{b'bv}_{ims}$. In both cases, the diagram
amplitude acquires the factor $g_1$. It is then obvious that all
possible diagram structures admissible by the Feynman rules can be
reproduced by the powers of the grammar matrix, $\mathbb{G}^n$,
$n\geq 1$. Here $n$ is the number of interaction vertices
appearing in the diagram (excepting the ultra-local vertices
$\widetilde{\bf v'}^m(x_0,t_0)$ and $\widetilde{\bf
b'}^m(x'_0,t'_0)$ and the starting vertex related to the point
$(x,t)$); it can be interpreted as the number of bifurcations of
moments in the certain magnetohydrodynamic process. The entries of
$\mathbb{G}^n$ are the monomials, the products of powers of the
coupling constants $g_{1,2,3,4}$.

The use of arguments given in Appendix \ref{Append:B} helps to verify that
the multiplicative factors for Feynman diagrams of the Green
 functions $G_v$ and $G_b$ contain powers of  $\mathcal{G}_v
 \equiv g_1^2g_2g_3g_4$, $\mathcal{G}_b \equiv g_2g^2_3g^2_4$
 (see Tab.~1). Namely these combinations of coupling constants
 play the role of the expansion parameters in the diagram series
  (\ref{diagrams}) for $G_v$ and $G_b$.

\vspace{0.5cm}
\leftline{\small \bf Table 1: The multiplicative factors of Feynman diagrams in MHD}
\vspace{0.3cm}
\begin{tabular}{||l||c|c||}
\hline\hline
 \# of bifurcations & $G_v$ & $G_b$ \\ \hline\hline
 $n\,=\,2k$, $k\geq 0$&
$\begin{array}{l}
g_1\cdot\mathcal{G}_v^k, \\
g_2\cdot\mathcal{G}_v^k,
\end{array}$ & $\begin{array}{l}
g_3\cdot\mathcal{G}_b^k, \\
g_4\cdot\mathcal{G}_b^k,
\end{array}$ \\ \hline
  $n\,=\,2k+1$, $k\geq 0$ &
$\begin{array}{l}
g_1g_2g_4\cdot\mathcal{G}_v^k, \\
g_1g_2g_3\cdot\mathcal{G}_v^k, \\
g^2_1g_2\cdot\mathcal{G}_v^k,
\end{array}$
 &
$\begin{array}{l}
g^2_3g_4\cdot\mathcal{G}_b^k, \\
g^2_4g_3\cdot\mathcal{G}_b^k, \\
g_2g_3g_4\cdot\mathcal{G}_b^k,
\end{array}$\\
 \hline\hline
\end{tabular}
\vspace{0.5cm}

Consequently, if in the equation $k_1+k_2+k_3+k_4=N$ we assume that
\begin{enumerate}
\item
either
$k_1=2k+1,$ $k_2=k_3=k_4=k$, or $k_1=2k,$ $k_2=k+1,$ $k_3=k_4=k$,
 then the diagram series (\ref{diagrams}) reproduces the
Feynman graphs for the Green function of fluid velocity $G_v$
containing an {\it even} number of bifurcation of moments;
\item
either
$k_1=2k+1,$ $k_2=k_4=k+1,$ $k_3=k$, or $k_1=2k+1,$ $k_2=k_3=k+1,$ $k_4=k$, or eventually
$k_1=2k+2,$ $k_2=k+1,$ $k_3=k_4=k$, then the diagram series (\ref{diagrams})
 reproduces the
Feynman graphs for the Green function  $G_v$ containing an {\it odd} number
of bifurcation of moments.
\end{enumerate}
Diagrams for the Green function of magnetic field $G_b$ do not contain the
 vertex responsible for the hydrodynamical interaction $V^{v'vv}_{ims}$,
 and therefore $k_1=0$.
Then, if we suppose that
\begin{enumerate}
\item
 either
 $k_2=k$, $k_3=2k+1,$ $k_4=2k$, or $k_2=k,$ $k_3=2k,$ $k_4=2k+1$, then the
  diagram series (\ref{diagrams}) reproduces the
Feynman graphs for  $G_b$ containing an {\it even} number of bifurcation
of moments;
\item
either
$k_2=k,$ $k_3=2k+2$, $k_4=2k+1,$ or  $k_2=k$, $k_3=2k+1,$ $k_4=2k+2$, or
 eventually
 $k_2=k+1,$ $k_3=2k+1$, $k_4=2k+1$, then the diagram series (\ref{diagrams})
  reproduces the
Feynman graphs for the Green function  $G_b$ containing an {\it odd} number
of bifurcation of moments.
\end{enumerate}
We have suggested everywhere that $k\in\mathbb{Z}_+$.

\section{The large order asymptotic behavior for the
MHD Green functions. Instanton approach and Borel summation }
\noindent

It is important to note that we do not know {\it apriori}
whether the coupling constants $g_{1,2,3,4}$
are small or large.
To get the information
on the convergence of asymptotic series (\ref{diagrams}), we
should study the asymptotic
behavior of large order coefficients $G^{(k_1,k_2,k_3,k_4)}_{\phi}$ ,
 $k_1+k_2+k_3+k_4=N,$ in the diagram
series as $N\to \infty,$ by the asymptotic calculation of the Cauchy
integral,
\begin{equation}
\label{Cauchy}
\begin{array}{lcl}
G^{(k_1,k_2,k_3,k_4)}_{\phi}&=&\frac 1{(2\pi)^4}\,
\frac 1{k_1!k_2!k_3!k_4!}\,\oint\,\prod_{i=1}^4\,\frac{dg_i}{g_i}\\
& \times & G_\phi\left(x,t;x_0,t_0;x'_0,t'_0\right)\,
\exp\left(-\sum_{i=1}^4\,k_i\,\log g_i\right),
\end{array}
\end{equation}
in which $G_\phi\left(x,t;x_0,t_0;x'_0,t'_0\right)$ is the functional
integral (\ref{G_v}). The contour of integration in the multiple integral
(\ref{Cauchy}) embraces
the points $\left\{g_i =0\right\}$, $i=1,\ldots 4,$ in the complex plane.

We estimate the functional integral (\ref{Cauchy}) by the steepest
descent method supposing that $\sum_{i=1}^4\,k_i=N$ is large (the instanton method).
In so far, the instanton approach has been
applied to various problems of stochastic dynamics, see \cite{Gurarie}-\cite{Honkonen}.
In contrast to all previous studies, in the MHD system we do not have one coupling
constant, but four. In the previous section, we have demonstrated that by applying
 the additional conditions for the integers $k_i$ to the equation
 $\sum_{i=1}^4\,k_i=N$ in the series (\ref{diagrams}), we can derive
 the diagram expansions for the different Green functions being in the
 diverse statistical regimes (characterized by the even and odd number
 of momentum bifurcations respectively).

Following the traditional instanton analysis, we perform the
uniform rescaling of variables in the action functional
(\ref{action}) in order to extract their dependence upon $N$,
\begin{equation}
\label{rescaling}
\begin{array}{l}
\widetilde{\phi} \,\to\, \widetilde{\phi}\,\sqrt {N}, \quad \widetilde{\phi\,'}\, \to \, \widetilde{\phi\,'} \,\sqrt {N},\\
 g_i \,\to\, g_i\,/N,\quad \nu \to \nu\,N,\quad \lambda \to \lambda\,
N,\quad x\to x\, \sqrt{N}.
\end{array}
\end{equation}
This keeps the action functional (\ref{action}) unchanged, thus
each term acquires the multiplier $N$ and then formally gets the
same order, as the $\log g_i$ in (\ref{Cauchy}) independently of
the type of interaction $i=1,\ldots 4.$ The corresponding
Jacobians from the numerator and the denominator of (\ref{G_v})
cancel. The saddle point equations are
\begin{equation}
\label{saddle_point_eqs}
\begin{array}{l}
\dot{\widetilde{v_i}} + g_1\,\nu\left(\widetilde{v}\,\partial\right)\,\widetilde{v_i}-
g_2\,\nu\left(\widetilde{b}\,\partial\right)\,\widetilde{b_i}-\nu\,\Delta\widetilde{v_i}\,=
\,\delta\left(x-x_0\right)\,\delta\left(t-t_0\right),\\
\dot{\widetilde{b_i}} + g_3\,\lambda\left(\widetilde{v}\,\partial\right)\,\widetilde{b_i}-
g_4\,\lambda\left(\widetilde{b}\,\partial\right)\,\widetilde{v_i}-\lambda\,\Delta\widetilde{b_i}\,=
\,\delta\left(x-x^{\,'}_0\right)\,\delta\left(t-t^{\,'}_0\right),\\
\dot{\widetilde{v^{\,'}_i}} + g_1\,\nu\left(\widetilde{v}\,\partial\right)\,\widetilde{v^{\,'}_i} - g_3\,\lambda\left(\widetilde{b^{\,'} }\partial\right)\,\widetilde{b_i} + \nu\,\Delta\widetilde{v^{\,'}_i}\,=\,0, \\
\dot{\widetilde{b^{\,'}_i}} + g_2\,\nu \left(\widetilde{v^{\,'}}\partial\right)\,\widetilde{b_i} +
g_3\,\lambda \left(\widetilde{b^{\,'}}\partial\right)\,\widetilde{v_i} +
\lambda\,\Delta\widetilde{b^{\,'}_i}\,=\,0, \\
\nu\,\widetilde{v_i}\left(\widetilde{v}\,\partial\right)\,\widetilde{v^{\,'}_i}\,=\,g^{-1}_1, \quad \nu\, \widetilde{v^{\,'}_i} \left(\widetilde{b}\,\partial\right)\,\widetilde{b_i}\,=\,g^{-1}_2,\\
\lambda\,\widetilde{b^{\,'}_i}\left(\widetilde{v}\,\partial\right)\,\widetilde{b_i}\,=\,g^{-1}_3,\quad \lambda\, \widetilde{b^{\,'}_i} \left(\widetilde{b}\,\partial\right)\,\widetilde{v_i}\,=\,g^{-1}_4.\\
\end{array}
\end{equation}
The first two equations in (\ref{saddle_point_eqs}) recover the original Cauchy problem
for the MHD equations. The next two equations (\#3 and \#4) occur
within the framework of the stochastic approach since we describe the
 microscopic dynamics in the MHD system as Brownian motion. The equations
  for the auxiliary fields are characterized by the negative viscosity and
   resistivity, and therefore
$\widetilde{{\bf v}^{\,'}}\left(t > t_0\right) = \widetilde{{\bf b}^{\,'}}\left(t > t^{\, '}_0\right) = 0$. The last four equations in (\ref{saddle_point_eqs}) determine the saddle-point values of the coupling constnts that allows to exclude
the interaction terms from the previous saddle-point equations and reduce
the system (\ref{saddle_point_eqs}) to
\begin{equation}
\label{saddle_point_eqs2}
\begin{array}{l}
\widetilde{{\bf v}^{\,'}}\,K_v\, \widetilde{\bf v}\,=\, \widetilde{{\bf v}^{\,'}}\, \,\delta\left(x-x_0\right)\,\delta\left(t-t_0\right),\\
\widetilde{{\bf b}^{\,'}} \,K_b\,\widetilde{\bf b}\,=\,\widetilde{{\bf b}^{\,'}} \, \delta\left(x-x^{\,'}_0\right)\,\delta\left(t-t^{\,'}_0\right),\\
\widetilde{\bf v}\,K^*_v\,\widetilde{{\bf v}^{\,'}} \,=\,1,\\
\widetilde{\bf b}\,K^*_b\,\widetilde{{\bf b}^{\,'}}\,=\, 1,
\end{array}
\end{equation}
in which we have introduced the diffusion kernels, $K_v\,=\,-i\omega+\nu\,p^2$
and $K_b\,=\,-i\omega+\lambda\,p^2$ (in the $(\omega,p)$ Fourier space) and $K^*_v$, $K^*_b$
are their Hermit conjugated forms.

Bifurcations of vortexes arisen due to the nonlinear interactions in the MHD
 system do not conclude
into a critical regime, and therefore the time spectrum in the
nonlinear model is the same as for the free diffusion equations,
$T\,\propto\, L^2$, that is the reason for the branching processes
are Poisson distributed with the characteristic times
$1/\left(\nu\,k^2\right)$ and  $1/\left(\lambda\,k^2\right)$. One
can see that the saddle-point configurations
$\left\{\widetilde{{\bf v}^{\,'}},\widetilde{{\bf b}^{\,'}},
\widetilde{\bf v},\widetilde{\bf b} \right\}$ which satisfy
(\ref{saddle_point_eqs2}) should be independent of the Poisson
branching processes, and therefore, the solutions could exist
before bifurcations start that is as $t\to \min\left(t_0,t^{\,
'}_0\right)$. With the use of the power model for the Dirac delta
function,
\[
\delta \left( {x} \right) = {\mathop {\lim} \limits_{\varepsilon
\to 0} }{\frac{{\varepsilon} }{{\pi \left( {x^{2} + \varepsilon
^{2}} \right)}}}{\rm ,}
\]
one can find that in the limit $t\to \min\left(t_0,t^{\, '}_0\right)$,
 (\ref{saddle_point_eqs2}) is satisfied
by the following radially symmetric solutions ($\phi\equiv \{v,b\}$,
 $\phi^{\,'}\equiv \{v^{\,'},b^{\,'}\}$, $\chi\equiv \{\nu,\lambda\}$, $t_0\equiv\{t_0,t'_0\}$),
\begin{equation}
\label{29}
\begin{array}{l}
  \phi\left( {{\rm {\bf r}},t} \right) = \sqrt {\left( {{\rm {\bf r}} - {\rm {\bf
r}}_{0}}  \right)^{2} + \left( {t - t_{0}}  \right)^{2}} ,\quad {\rm}
{\rm }{\phi}'\left( {{\rm {\bf r}},t} \right) = {\frac{{H \left(
{t_{0} - t} \right)}}{{t - t_{0}} }}\phi\left( {{\rm {\bf r}},t}
\right), \\
\phi\left( {{\rm {\bf r}},t} \right) = {\frac{{H \left( {t -
t_{0}} \right)}}{{\pi} }}\arctan \left( {{\frac{{{\left| {{\rm
{\bf r}} - {\rm {\bf r}}_{0}}  \right|}}}{{2\chi \left( {t - t_{0}}
\right)}}}} \right), \quad {\rm }{\rm} {\phi}'\left( {{\rm {\bf r}},t}
\right) = - {\frac{{\pi} }{{2\chi  }}}{\frac{{\left( {4\chi
^{2}\left( {t - t_{0}}  \right)^{2} + \left( {{\rm {\bf r}} - {\rm
{\bf r}}_{0}}  \right)^{2}} \right)}}{{{\left| {{\rm {\bf r}} -
{\rm {\bf r}}_{0}}  \right|}}}}, \\
\phi\left( {{\rm {\bf r}},t} \right) = {\frac{{H\left( {t -
t_{0}} \right)}}{{\pi} }}\arctan \left( {{\frac{{2\chi \left( {t -
t_{0}} \right)}}{{{\left| {{\rm {\bf r}} - {\rm {\bf r}}_{0}}
\right|}}}}} \right), \quad {\rm} {\rm} {\phi}'\left( {{\rm {\bf r}},t}
\right) = {\frac{{\pi }}{{2\chi} }}{\frac{{\left( {4\chi ^{2}\left(
{t - t_{0}}  \right)^{2} + \left( {{\rm {\bf r}} - {\rm {\bf
r}}_{0}}  \right)^{2}} \right)}}{{{\left| {{\rm {\bf r}} - {\rm
{\bf r}}_{0}}  \right|}}}},
\end{array}
\end{equation}
In the first solution (\ref{29}), the auxiliary fields $v'$ and $b'$ have poles as
either $t \to t_0$ or $t\to t'_0$, and then it follows from (\ref{saddle_point_eqs})
 that $g_{1}^*=g_2^*=0$ and $g_{3}^*=g_4^*=0$, the point which
definitely lays inside the integration contour in (\ref{Cauchy}).
 In contrast to it, in the last equation in (\ref{29}), $v\to 0 $ as $t\to t_0$ and
$b\to 0$ as $t\to t'_0$ respectively,
and therefore $g_{1,2,3,4}^*\to\infty$ that is definitely outside the
integration contour. There is a subtle point in
Eq.~(\ref{29}) concerning the second solution since
$v(r,t=t_0)=b(r,t=t'_0)=H(0)/2,$ and the position of the saddle-point
configuration charges $g_{1,2,3,4}^*$ depends upon the conventional value
for the Heaviside function of zero argument,
$H(0).$ While estimating the functional Jacobian,
we had assumed following the standard convention  \cite{Adzhemyan:1999}
 that $H(0)=0$. Then, one can easily verify that in this case we also
have  $g_{1,2,3,4}^*\to
\infty$, so that being interested in the large order asymptotic
behavior of the Green functions in MHD we do not need to take the second and
the third solutions into account. Even if one takes $H(0)\ne 0$,
and then $0<g_{1,2,3,4}^*<\infty,$ it is always possible to deform the
integration contour in (\ref{Cauchy}) in such a way
to avoid $g_{1,2,3,4}^*$ to be encircled.
Therefore, the first solution in (\ref{29}) is the only one we need, and
substituting it into (\ref{saddle_point_eqs}), we can obtain the microscopic
 power models for the coupling constants,
\begin{equation}
\label{power_g}
g_{1,2}^*\equiv \lim_{\delta t \to 0} g_{1,2} \simeq \frac{\delta t}{\nu\, \delta
r^{2}},\quad g_{3,4}^*\equiv \lim_{\delta t \to 0} g_{3,4} \simeq \frac{\delta t}{\lambda\, \delta
r^{2}}.
\end{equation}
Fields $\phi$, $\phi',$ and the coupling constants $g_{1,2,3,4}$
fluctuate around their saddle-point values $\phi_*,$ $\phi'_*$,
and $g_{1,2,3,4}^*$. By means of the standard shift of variables,
$\phi=\phi_*+\delta \phi,$ $\delta \phi'=\phi'_*+\delta \phi',$
$g_{1,2,3,4}=g_{1,2,3,4}^*+\delta g_{1,2,3,4}$, one makes them
fluctuate around zero, so that $\delta \phi (\infty)\,=\,0$,
$\delta \phi'(\infty)\,=\,0$. Moreover, if we assume that the MHD
system is isotropic (i.e., there is no the global bias of the
magnetic fields and the conductive
 fluid is isotropic), then
 $\mathrm{tr}\, \delta \phi\, =\,\mathrm{tr}\, \delta \phi' =0$,
and therefore all fluctuations
posses the central symmetry, $\delta \phi\, =\,\delta \phi ({\bf r},t)$, $\delta
\phi'\,=\,\delta \phi' ({\bf r},t)$, the same as the saddle point configuration.

The contours of integrations over the variables $\delta g_{1,2,3,4}$ now passes
through the origin, and are directed there oppositely to the
imaginary axis. The integrals over $\delta g_{1,2,3,4}$ are conducted now on the
rectilinear contours in complex planes $(i\infty, -i\infty)$ (in
accordance to the standard transformation of contours in the
steepest descent method). At the turn of the integration contours
$\delta g_{1,2,3,4} \to -i\,\delta g_{1,2,3,4}$,
where the multiplier $(-i)$ appears so that
the result $G_{\phi}^{(k_1,k_2,k_3,k_4)}$ is always real.
The contributions to the Cauchy
integral (\ref{Cauchy}) comes from the poles
$\delta g_{1,2,3,4}\, =\,-g_{1,2,3,4}^*$ and tends to zero as
$t\to \min\left(t_0,t'_0\right).$
The values of the functional integrals on the saddle
point configurations are proportional to the entire volume of the
functional integration, they cancel in the numerator and
denominator simultaneously. While calculating the fluctuation
integral (\ref{G_v}), we take into account that the first order contributions
in $\delta \phi$ and $\delta \phi'$ are absent because of the
saddle-point condition. We also neglect the high-order
interactions between fluctuations, $O(\delta \phi^3),$ $O(\delta
\phi^4),$ \textit{etc.} to arrive at the Gaussian functional
integrals for the Green functions $G_v$ and $G_b$,
\begin{equation}
\label{functionals}
\begin{array}{lcl}
G_v^{(N)}(r,\delta t)&
\simeq_{\delta t\rightarrow 0}&\left(\mathcal{G}_v^*\right)^{\frac{N}{2}}\,\frac{ N^{N+1/2}}{2\pi} \\
& \times&
 \int \! \!\int \,\mathcal{D}\,\delta \phi\, \mathcal{D}\,\delta \phi'\,
\exp \,-\frac N2\, \mathrm{tr}\left[ \delta \phi' K_\phi \delta \phi
+\frac{\nu}{\lambda}\,\delta \phi\left( \frac 1r \partial_r\right)\delta \phi \right],\\
G_b^{(N)}(r,\delta t)&
\simeq_{\delta t\rightarrow 0}&\left(\mathcal{G}_b^*\right)^{\frac{N}{2}}\,\frac{ N^{N+1/2}}{2\pi}\\
& \times&
 \int \! \!\int \,\mathcal{D}\,\delta \phi\, \mathcal{D}\,\delta \phi'\,
\exp \,-\frac N2\, \mathrm{tr}\left[ \delta \phi' K_\phi \delta \phi
+\frac{\lambda}{\nu}\,\delta \phi\left( \frac 1r\, \partial_r\right)\delta \phi \right],
\end{array}
\end{equation}
in which $\mathcal{G}_v$ and $\mathcal{G}_b$ are the
the expansion parameters in the diagram series (\ref{diagrams}) for $G_v$ and $G_b$
introduced in Sec.~\ref{Sec:factors} in concern with Tab.1.
Performing the usual rescaling of fluctuation fields
$$
\delta \phi \to \delta \phi /\sqrt{N}, \quad \delta \phi' \to \delta
\phi'/\sqrt{N},
$$
we compute the Gaussian integral, with respect to $\delta \phi$
first, and then the resulting Gaussian integral over the fluctuation of the auxiliary fields $\delta \phi'$,
\begin{equation}
\label{33}
\begin{array}{lcl}
G_v^{(N)}(r,\delta t)&\simeq_{\delta t\rightarrow 0}N^{N-1/2}&
\exp \left(-\frac N2\, \log \left(\mathcal{G}_v^*\right) \right)\,\det K_\phi^{-1}\,
\left(1+O\left(\frac{1}{N}\right)\right), \\
G_b^{(N)}(r,\delta t)&\simeq_{\delta t\rightarrow 0}N^{N-1/2}&
\exp \left(-\frac N2\, \log \left(\mathcal{G}_b^*\right) \right)\,\det K_\phi^{-1}\,
\left(1+O\left(\frac{1}{N}\right)\right). \\
\end{array}
\end{equation}
 The kernels of the operators $K_{\phi}^{-1}$ are the Green function
of the linear diffusion equations. Using the Stirling's
formula, one can check that the coefficients $G_\phi^{(N)}$ of the
asymptotic series (\ref{diagrams}) demonstrate the factorial growth (like
in the most of quantum field theory models):
\begin{equation}
\label{34}
\begin{array}{lcl}
G_v^{(N)}(r,\delta t) &\simeq_{\delta t\rightarrow 0}  & \frac{N!}{2\pi
N}\exp N\left(1-\frac 12\,\log \left(\mathcal{G}_v^*\right)\right)\det K_\phi^{-1}
\left(1+O\left(\frac{1}{N}\right)\right),\\
G_b^{(N)}(r,\delta t) &\simeq_{\delta t\rightarrow 0}  & \frac{N!}{2\pi
N}\exp N\left(1-\frac 12\,\log \left(\mathcal{G}_b^*\right)\right)\det K_\phi^{-1}
\left(1+O\left(\frac{1}{N}\right)\right).\\
\end{array}
\end{equation}
Therefore, the asymptotic series (\ref{diagrams}) can be summed by means
of Borel's procedure. It consists of the following transformation
of series (\ref{diagrams})
\begin{equation}
\label{35}
\begin{array}{lcl}
\sum_{N}G_\phi^{(N)}\mathcal{G}_\phi^N& = & \sum_{N}\,\Gamma(N+1)\,\widetilde{G}_\phi^{(N)}\,\mathcal{G}_\phi^N\\
& =& \sum_{N}\,\int_{0}^{\infty}\,
d\tau \,\, \widetilde{G}_\phi^{(N)} \left(\tau\,\mathcal{G}_\phi\right)^{N} e^{ -\tau},\\
\end{array}
\end{equation}
where $ \widetilde{G}_\phi^{(N)} =G_\phi^{(N)}/\Gamma(N+1)$ are the new
expansion coefficients which do not exhibit the factorial growth.

It is traditional, while performing the Borel summation,
to change the orders of summation and integration in (\ref{35})
\begin{equation}
\label{36}
\sum_{N}\,\int_{0}^{\infty}\, d\tau \,\, \widetilde{G}_\phi^{(N)}\left(\mathcal{G}_\phi\tau\right)^{N}
e^{ -\tau}\, =\, \int_{0}^{\infty}\, d\tau \,\, e^{ -\tau}\,\sum_{N}
\widetilde{G}_\phi^{(N)}(\mathcal{G}_\phi\tau)^{N}.
\end{equation}
The we sum over $N$ in the r.h.s. of (\ref{36}),
\[
\frac{\det K_\phi^{-1}}{2\pi}\,\int_{0}^{\infty}\,
 d\tau \,\,e^{ -\tau}\,\sum_{N=1}^{\infty}\,
 \frac{\left(\mathcal{G}_\phi\tau\right)^{N}}{N}\,e^{N\left( 1- \log
\mathcal{G}_\phi^*\right)},
\]
 and obtain
\begin{equation}\label{37}
 =  -\frac{\det K_\phi^{-1}}{2\pi}\,\int_{0}^{\infty}\,
 d\tau \,\,e^{ -\tau}\, \log\left(1-\tau\frac{\mathcal{G}_\phi}{\mathcal{G}^*_\phi} \right).
\end{equation}
The integration of (\ref{37}) over $\tau$ gives us
\begin{equation}
\label{38}
  G_\phi\,\simeq _{\delta t \to 0}\,\det K_\phi^{-1}\left(
  1+\frac{1}{2\pi}
  \mathrm{Ei}\left( \frac{\mathcal{G}^*_\phi}{\mathcal{G}_\phi}\right)\,
 \exp \left(-\frac{\mathcal{G}^*_\phi}{\mathcal{G}_\phi}\right)\right)
\end{equation}
where $\mathrm{Ei}(x)$ is the exponential integral defined as $\mathrm{Ei}(x) = -\int_{-x}^\infty \, y^{-1}\,e^{-y}\, dy$ understood in terms of the Cauchy principal value at $y=0$.
Then we can use the microscopic models (\ref{power_g}) and recall that by definition $g_{1,2}=1/\nu$ and $g_{3,4}=1/\lambda$ in order to estimate the both
ratios $\mathcal{G}^*_v/\mathcal{G}_v$ and $\mathcal{G}^*_b/{\mathcal{G}_b}$ in (\ref{38}) as
\begin{equation}
\label{ratios}
\frac{\mathcal{G}^*_\phi}{\mathcal{G}_\phi}\,\simeq_{\delta t \to 0}\,
\frac{\delta t ^{5/2}}{\delta r^5}.
\end{equation}
In such a simplified model, the non-Maxwellian corrections to the
distribution functions arisen due to the kinetic effects are
accounted by the new distribution function,
\begin{equation}
\label{39}
G_\phi(\delta t,\delta r)\,\simeq _{\delta t \to 0} \,\frac{\exp \left(
-\frac{\delta r^2}{4\chi (\delta t )}\right)} {\left( 4\pi \chi (\delta t)\right)^{d/2}}\left[
  1+\frac{1}{2\pi}
  \mathrm{Ei}\left( \frac{\left(\delta t\right)^{5/2}}{\left( \delta r
   \right)^{5}}\right) \exp \left(-\frac{\left(\delta t\right)^{5/2}}{\left( \delta r
   \right)^{5}}\right)\right].
\end{equation}

\section{Discussion and Conclusion}
\noindent

Speaking rigorously, the classical MHD equations cannot be applied if
the time scale of collisions is comparable or longer  than the
other characteristic times in plasmas. The particle distributions
are far from being of the Maxwellian shape
and the certain kinetic models giving an insight into the collision
 statistics have to be taken into account.
The simplest such model which allows accounting the kinetic
effects being nevertheless completely in the framework of the
classical approach based on the MHD equations is suggested in the
present paper.

We investigate a simple generalization of the MHD model (\ref{MHD})
modelling fluctuations of the configurations $\{{\bf v},{\bf b}\}$ considered as
the stochastic fields (or the trial trajectories of the MHD system) for which the
classical MHD (\ref{MHD}) plays the role of the mean field equations.

The essential point of our approach
is that we have used the field theory formulation of the dynamics
which allowed us the implementation of various powerful
technics borrowed from the quantum field theory. In particular, with
the use of the instanton technique and Borel's summation, we
have computed the asymptotic series for the Green functions
accounting for the kinetic effects as the corrections to the unperturbed
 diffusion kernel
describing the pure relaxation dynamics.
Similarly to the most of quantum field theory models, the high
order contributions into the Green functions exhibit a factorial
growth.
\begin{figure}[hp]
 \noindent
 \begin{minipage}[b]{.36\linewidth}
 \begin{center}
 \epsfig{file=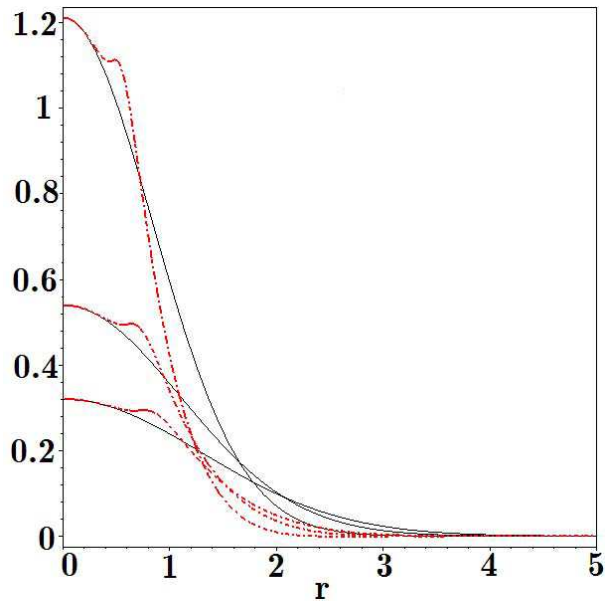,  angle= 0,width =8cm, height =8cm}
 \end{center}
\end{minipage}
\label{Fig5}
 \caption{ The profiles of standard diffusion kernel $G(\delta r)$
(the solid line) calculated at at several consequent time
steps $t>0$ for $\nu=0.2$. The
dash-dot lines present the asymptotic kernel (\ref{39}) (as $t\to t_0$)
accounting for the perturbation due to the kinetic effects modelled by Brownian motion.}
\end{figure}

\begin{figure}[hp]
 \noindent
 \begin{minipage}[b]{.36\linewidth}
 \begin{center}
 \epsfig{file=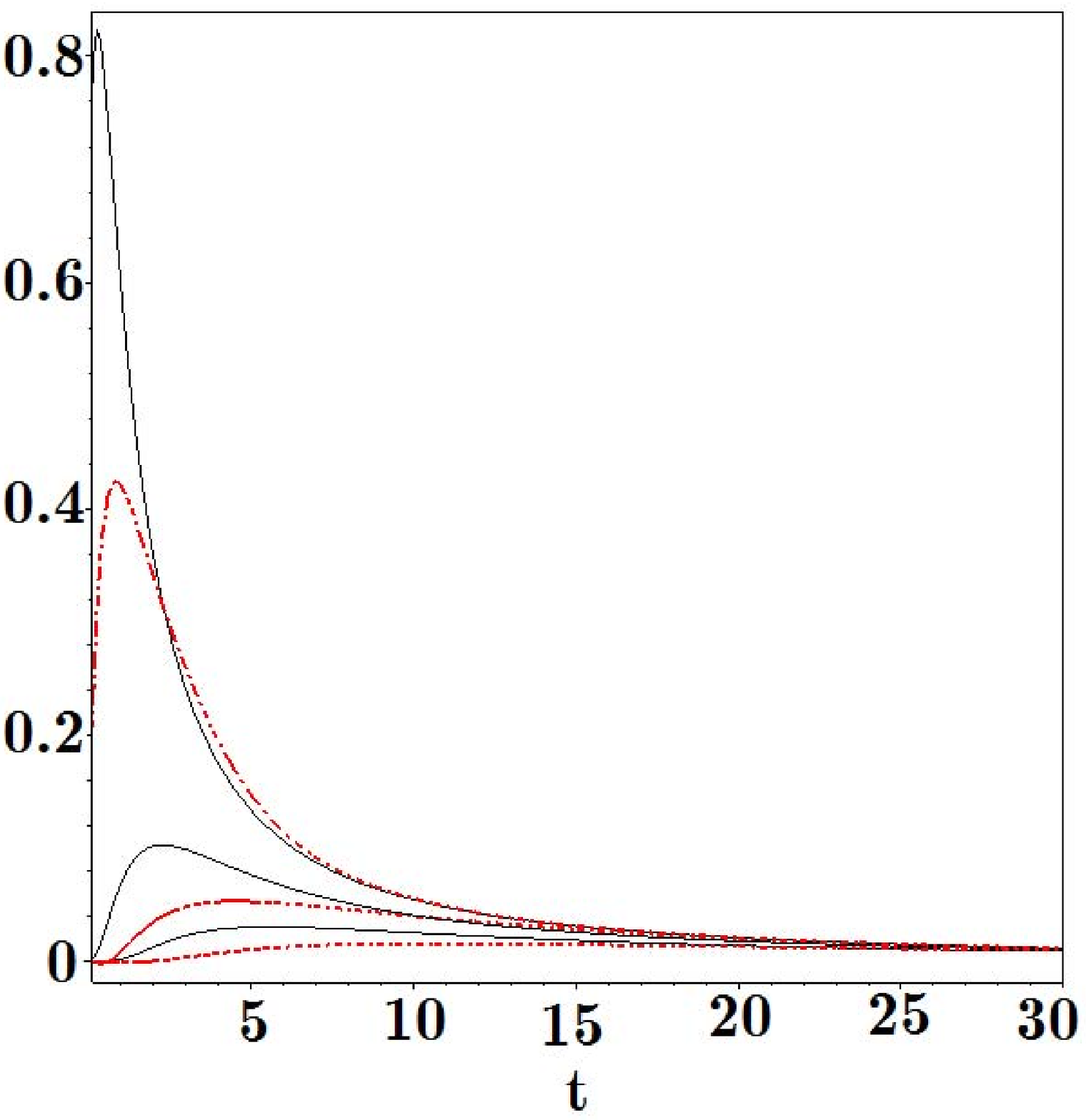,  angle= 0,width =8cm, height =8cm}
 \end{center}
\end{minipage}
\label{Fig6}
 \caption{ The profiles of standard diffusion kernel $G(\delta t)$
(the solid line) calculated at several distant points for $\nu=0.2$. The
dash-dot lines present the asymptotic kernel (\ref{39}) (as $t\to t_0$)
accounting for  the perturbation due to the kinetic effects modelled by Brownian motion.}
\end{figure}

It is interesting to compare the perturbed diffusion kernel
(\ref{39}) in the 3D space, $d=3,$ with the standard diffusion
kernel depicted by a Gaussian curve. In
Figs.~\ref{Fig5},\ref{Fig6} we have sketched the profiles of
standard diffusion kernel (the solid lines) calculated at at
several consequent time steps for $\nu=0.2$ together with the
perturbed kernel profiles given by (\ref{39}).

It is clearly seen that the essential corrections to the standard diffusion kernel are arisen in short times and small scales, while they are negligible in long times large scales.

\section*{Acknowledgment}
\label{Acknowledgment}
\noindent

This work has been started while D.V. was a guest researcher at the {\it Centre de Physique Th\'{e}orique}, CNRS - UMR 6207.
Now D.V. has got a support from the Volkswagen Foundation (Germany)
in the framework of the project "{\it Network formation rules, random
set graphs and generalized epidemic processes}" (Contract no Az.:
I/82 418).

\appendix
\section{Analytical expressions for Feynman diagrams}

Below, we write down the analytical expressions (in
$d$-dimensional Fourier space $({\bf k},t)$) correspondent to
diagrams shown in Figs.~(\ref{Fig1},\ref{Fig2}).
\begin{equation}
\label{A1}
\begin{array}{l}
G_v({\bf k},t-t_0;{\bf k'},t-t'_0)= v_0({\bf k})R^{v'v}({\bf k},t-t_0)\\
+\int^\infty_0 dt'V^{v'vv}({\bf k})R^{v'v}({\bf k},t-t') \int\frac{d{\bf q}}{(2\pi)^d}  R^{v'v}({\bf q},t'-t_0)v_0({\bf q})R^{v'v}({\bf k-q},t'-t_0)\\
v_0({\bf k-q}) \\
+\int^\infty_0 dt'V^{v'vv}({\bf k})R^{v'v}({\bf k},t-t') \int\frac{d{\bf q}}{(2\pi)^d}R^{v'v}({\bf k-q},t'-t_0)v_0({\bf k-q})  \\
\int dt'' V^{v'vv}({\bf q}) R^{v'v}({\bf q},t'-t'') \int\frac{d{\bf p}}{(2\pi)^d}R^{v'v}({\bf p},t'-t_0) v_0({\bf p})R^{v'v}({\bf q-p},t'-t_0)\\
v_0({\bf q-p}) \\
+\int^\infty_0 dt'V^{v'bb}({\bf k})R^{v'v}({\bf k},t-t') \int\frac{d{\bf q}}{(2\pi)^d}  R^{b'b}({\bf q},t'-t'_0)b_0({\bf q})\\
R^{b'b}({\bf k-q},t'-t'_0)b_0({\bf k-q}) \\
+\int^\infty_0 dt''V^{v'vv}({\bf k})R^{v'v}({\bf k},t''-t)\int\frac{d{\bf k-p}}{(2\pi)^d} v_0({\bf k- p})R^{v'v}({\bf k-p},t''-t_0)\\
\int^\infty_0 dt'V^{v'bb}({\bf p})R^{v'v}({\bf p},t'-t'') \int\frac{d{\bf q}}{(2\pi)^d}  R^{b'b}({\bf q},t'-t'_0) b_0({\bf p})
b_0({\bf p-q})\\
R^{b'b}({\bf p-q},t'-t'_0) \\
+\int^\infty_0 dt''V^{v'vv}({\bf k})R^{v'v}({\bf k},t-t'')\int\frac{d{\bf p}}{(2\pi)^d} R^{v'v}({\bf p},t'-t'') R^{v'v}({\bf k-p},t'''-t'') \\
V^{v'bb}({\bf p})V^{v'vv}({\bf k-p})\int\frac{d{\bf q}}{(2\pi)^d}  R^{b'b}({\bf q},t'-t'_0) b_0({\bf p})
b_0({\bf p-q})\int\frac{d{\bf s}}{(2\pi)^d} R^{v'v}({\bf s},t'''-t_0) )\\
R^{v'v}({\bf k-p-s},t'''-t_0) v_0({\bf s})v_0({\bf k-p-s})
\\
+\int^\infty_0 dt''V^{v'vv}({\bf k})R^{v'v}({\bf k},t-t'')\int^\infty_0 dt'''\int\frac{d{\bf p}}{(2\pi)^d} R^{b'b}({\bf p},t'''-t'')V^{b'bv}({\bf p})\\
V^{b'bv}({\bf k-p})
R^{b'b}({\bf k-p},t''-t')\int\frac{d{\bf q}}{(2\pi)^d}
 v_0({\bf q})v_0({\bf -q}) b_0({\bf q})b_0({\bf -q})R^{b'b}({\bf q},t'''-t'_0)\\
R^{v'v}({\bf q},t'-t_0)R^{v'v}({\bf -q},t'''-t_0)R^{b'b}({\bf -q},t'-t'_0)
\end{array}
\end{equation}

\begin{equation}
\label{A2}
\begin{array}{l}
G_b({\bf k},t-t_0;{\bf k'},t-t'_0)= b_0({\bf k'})R^{b'b}({\bf k'},t-t'_0)\\
+\int^\infty_0 dt'V^{b'vb}({\bf k'})R^{b'b}({\bf k'},t-t') \int\frac{d{\bf q}}{(2\pi)^d}R^{v'v}({\bf k'-q},t'-t'_0)b_0({\bf k-q})  \\
\int dt'' V^{v'bb}({\bf q}) R^{b'b}({\bf q},t'-t'') \int\frac{d{\bf p}}{(2\pi)^d}R^{b'b}({\bf p},t'-t'_0) b_0({\bf p})R^{b'b}({\bf q-p},t'-t'_0)\\
b_0({\bf q-p}) \\
+\int^\infty_0 dt'V^{b'vb}({\bf k'})R^{b'b}({\bf k'},t-t') \int\frac{d{\bf s}}{(2\pi)^d}v_0({\bf s})R^{v'v}({\bf k'-s},t'-t_0)\\
\int\frac{d{\bf q}}{(2\pi)^d}b_0({\bf q})R^{b'b}({\bf k'-q},t'-t'_0)\\
+ \int^\infty_0 dt'V^{b'vb}({\bf k'})R^{b'b}({\bf k'},t-t')\int^\infty_0 dt''
\int\frac{d{\bf s}}{(2\pi)^d}b_0({\bf s})R^{v'v}({\bf k'-s},t''-t'_0)
\\
\int\frac{d{\bf q}}{(2\pi)^d}v_0({\bf k'-s-q})v_0({\bf k'+s})R^{v'v}({\bf k'-s-q},t'-t_0)R^{v'v}({\bf k'+s},t''-t_0)
\\
V^{b'vb}({\bf k'-q-s})R^{b'b}({\bf k'-s-q},t'-t'')\\
+\int^\infty_0 dt''\int^\infty_0 dt'V^{b'vb}({\bf k'})R^{b'b}({\bf k'},t-t')  \\
\int\frac{d{\bf q}}{(2\pi)^d}V^{b'vb}({\bf k'-q-s})R^{b'b}({\bf k'-s-q},t'-t'')\int\frac{d{\bf q'}}{(2\pi)^d}\int\frac{d{\bf p}}{(2\pi)^d}
  R^{v'v}({\bf q},t'-t_0)v_0({\bf q'})
\\
v_0({\bf k'-q'})R^{v'v}({\bf k'-q'},t''-t_0)R^{v'v}({\bf q'},t''-t_0)V^{v'vv}({\bf k'-p})
\end{array}
\end{equation}

\section{The "grammar" matrix for the Feynman diagram technique}
\label{Append:B}

The "grammar" matrix $\mathbb{G}$ is the weighted  connectivity matrix, which
expresses the fact that  the tails $\phi$, $\phi'$ belonging to the interaction vertices $V^{v'vv}_{ims}$, $V^{v'bb}_{ims}$, and $V^{b'bv}_{ims}$ (being proportional to $g_1$, $g_2$, $g_3$, and $g_4$) can be connected to each other accordingly to the Feynmann rules only by means of the propagators (\ref{edges}).

\begin{center}
\begin{tabular}{||c||c|c|c|c|c|c|c|c|c|c|c|c|c||}
\hline
   &  \multicolumn{3}{l}{$g_1$} \vline& \multicolumn{3}{l}{$g_2$} \vline&  \multicolumn{3}{l}{$g_3$} \vline &  \multicolumn{3}{l}{$g_4$}\vline \\ \hline
 & $\bf v'$ & $\bf v$ & $\bf v$ & $\bf v'$ & $\bf b$ & $\bf b$ &  $\bf b'$ & $\bf v$ & $\bf b$ & $\bf b'$ & $\bf b$ & $\bf v$ \\ \hline\hline
  $\bf v'$ & 0 & $g_1$ & $g_1$ & 0 & 0 & 0 & 0 & $g_1$ & 0 & 0 & 0 & $g_1$ \\  \hline
  $\bf v$ & $g_1$ & 0 & 0 & $g_2$ & 0 & 0 & 0 & 0 & 0 & 0 & 0 & 0 \\   \hline
  $\bf v$ & $g_1$ & 0 & 0 & $g_1$ & 0 & 0 & 0 & 0 & 0 & 0 & 0 & 0 \\    \hline
  $\bf v'$& 0 & $g_2$ & $g_2$ & 0 & 0 & 0 & 0 & $g_2$ & 0 & 0 & 0 & $g_2$ \\    \hline
  $\bf b$ & 0 & 0 & 0 & 0 & 0 & 0 & $g_2$ & 0 & 0 & $g_2$ & 0 & 0 \\    \hline
  $\bf b$ & 0 & 0 & 0 & 0 & 0 & 0 & $g_2$ & 0 & 0 & $g_2$ & 0 & 0 \\   \hline
  $\bf b'$& 0 & 0 & 0 & 0 & $g_3$ & $g_3$ & 0 & 0 & $g_3$ & 0 & $g_3$ & 0 \\   \hline
  $\bf v$ & $g_3$ & 0 & 0 & $g_3$ & 0 & 0 & 0 & 0 & 0 & 0 & 0 & 0 \\    \hline
  $\bf b$ & 0 & 0 & 0 & 0 & 0 & 0 & $g_3$ & 0 & 0 & $g_3$ & 0 & 0 \\    \hline
  $\bf b'$& 0 & 0 & 0 & 0 & $g_4$ & $g_4$ & 0 & 0 & $g_4$ & 0 & $g_4$ & 0 \\   \hline
  $\bf b$ & 0 & 0 & 0 & 0 & 0 & 0 & $g_4$ & 0 & 0 & $g_4$ & 0 & 0 \\   \hline
  $\bf v$ & $g_4$ & 0 &0 & $g_4$ & 0 & 0 & 0 & 0 & 0 & 0 & 0 & 0 \\ \hline
\end{tabular}
\end{center}

The starting vertex in a diagram contributing into the Green function $G_v$ may be either
 $V^{v'vv}_{ims}$ or  $V^{v'bb}_{ims}$. In the first case, the multiplicative
  factor acquires the additional multiplier $g_1$, and it is $g_2$ if
   the second vertex is used as the starting one. In order to include
   the starting nodes into account, we introduce the vector
$\mathfrak{v}^\top=\left[g_1,0,0,g_2,0,0,0,0,0,0,0,0\right]$. Then,
 the multiplicative factors of the diagrams  for the Green function
  $G_v$ are written as following,
\begin{equation}
\label{diag2}
\left[\,
G_v\right]_{\mathrm{factor}}\,=\,\left\{
\,\mathbb{G}^n\mathfrak{v}\,\right\}_{n\,\geq\,0}.
\end{equation}
Similarly, we can find the multiplicative factors arising in the diagrams
for the Green function $G_b$ by
\begin{equation}
\label{diag3}
\left[\,
G_b\right]_{\mathrm{factor}}\,=\,\left\{
\,\mathbb{G}^n\mathfrak{b}\,\right\}_{n\,\geq\,0},
\end{equation}
in which the vector $\mathfrak{b}^\top=\left[0,0,0,0,0,0,g_3,0,0,g_4,0,0\right]$
expresses the fact that $V^{b'bv}_{ims}$ is the starting vertex for $G_b$.

\end{document}